\documentclass[aps,twocolumn,showpacs,superscriptaddress,reprint,prl]{revtex4-1}

\usepackage{amsthm}
\usepackage{amsfonts}
\usepackage{amsmath}
\usepackage{siunitx}
\usepackage{multirow}
\usepackage{graphicx}
\usepackage{caption}
\usepackage[usenames,dvipsnames,svgnames,table]{xcolor}
\usepackage{dcolumn}
\usepackage{bm}
\usepackage{gensymb}

\begin{document}

\title{Insights into ultrafast Ge-Te bond dynamics in a phase-change superlattice}

\author{Marco Malvestuto}
\email{marco.malvestuto@elettra.eu}
\affiliation{Elettra-Sincrotrone Trieste S.C.p.A. Strada Statale 14 - km 163.5 in AREA Science Park 34149 Basovizza, Trieste, Italy}
\author{Antonio Caretta}
\affiliation{Elettra-Sincrotrone Trieste S.C.p.A. Strada Statale 14 - km 163.5 in AREA Science Park 34149 Basovizza, Trieste, Italy}
\author{Barbara Casarin}
\affiliation{Department of Physics, University of Trieste, Via A. Valerio 2, 34127 Trieste, Italy}
\affiliation{Elettra-Sincrotrone Trieste S.C.p.A. Strada Statale 14 - km 163.5 in AREA Science Park 34149 Basovizza, Trieste, Italy}
\author{Federico Cilento}
\affiliation{Elettra-Sincrotrone Trieste S.C.p.A. Strada Statale 14 - km 163.5 in AREA Science Park 34149 Basovizza, Trieste, Italy}
\author{Martina Dell'Angela}
\affiliation{Elettra-Sincrotrone Trieste S.C.p.A. Strada Statale 14 - km 163.5 in AREA Science Park 34149 Basovizza, Trieste, Italy}
\author{Daniele Fausti}
\affiliation{Department of Physics, University of Trieste, Via A. Valerio 2, 34127 Trieste, Italy}
\author{Raffaella Calarco}
\affiliation{Paul-Drude-Institut f\"{u}r Festk\"{o}rperelektronik, Hausvogteiplatz 5-7, 10117 Berlin, Germany}
\author{Bart J. Kooi}
\affiliation{Zernike Institute for Advanced Materials, University of Groningen, Groningen, 9747 AG, The Netherlands}
\author{Enrico Varesi}
\affiliation{Micron Semiconductor Italia S.r.l., Via C. Olivetti, 2, 20864, Agrate Brianza, MB, Italy}
\author{John Robertson}
\affiliation{Engineering Department, Cambridge University, Cambridge CB2 1PZ, UK}
\author{Fulvio Parmigiani}
\affiliation{Department of Physics, University of Trieste, Via A. Valerio 2, 34127 Trieste, Italy}
\affiliation{Elettra-Sincrotrone Trieste S.C.p.A. Strada Statale 14 - km 163.5 in AREA Science Park 34149 Basovizza, Trieste, Italy}
\affiliation{International Faculty, University of Cologne, 50937 Cologne, Germany}



\begin{abstract}
A long-standing question for avant-grade data storage technology concerns the nature of the ultrafast photoinduced phase transformations in the wide class of chalcogenide phase-change materials (PCMs).
Overall, a comprehensive understanding of the microstructural evolution and the relevant kinetics mechanisms accompanying the out-of-equilibrium phases is still missing. Here, after overheating a phase-change chalcogenide superlattice by an ultrafast laser pulse, we indirectly track the lattice relaxation by time resolved X-ray absorption spectroscopy (tr-XAS) with a sub-ns time resolution.
The novel approach to the tr-XAS experimental results reported in this work provides an atomistic insight of the mechanism that takes place during the cooling process, meanwhile a first-principles model mimicking the microscopic distortions accounts for a straightforward representation of the observed dynamics. 
Finally, we envisage that our approach can be applied in future studies addressing the role of dynamical structural strain in phase-change materials.
\end{abstract}

\maketitle

\section{introduction}

 In these last years innovative fields for cutting-edge technologies based on novel engineered  materials have been disclosed by the understanding of the non-equilibrium optical control of the matter. 

For instance the comprehension of the non-equilibrium mechanisms is of paramount importance for exploiting the physical and optical properties of the phase-change materials (PCMs), nowadays used in optical data storage \cite{Wuttig:2007bs} and non-volatile electrical memories \cite{Lankhorst:2005cg}.

The key feature of these intriguing compounds is the large and steep change of the optical and electrical properties observed when comparing   the covalent bonded amorphous phase with the resonantly bonded crystalline phase. Interestingly, this scenario has been recently enriched by the chalcogenide superlattice’ (CSL), that are regarded as novel phase-change materials \cite{Tominaga:2008fh,Simpson:2011kh}, where the phase transition is between two crystalline structures, rather than amorphous to crystalline or vice-versa.

However, it is stimulating the fact that all these materials share common phase change properties, such as the switching time, the activation energy and the dielectric response, hence suggesting that a similar physics must govern the complex nature of their local atomic structure and configuration conditions.   


To shed light on these compelling mechanisms, some models, based mainly on thermal or electronic processes, have been proposed.\cite{Baker:2006br,Yu:2015jw,Sun:2006eb,Weinic:2007gq,sun:2006,Klein:2008kk,Caravati:2009hp,Lee:1972gx,Boer:1970hj,Makino:2012bq}

Conversely, other studies \cite{LeGallo:2016ga,Bruns:2009gj,Krebs:2009is,Kohary:2011bn,VazquezDiosdado:2012cv,Cao:2015bq} suggest that more complex mechanisms are governing the atomic dynamics at the base of the phase switching, where concomitant thermal and electronic aspects compete in a synergic feedback loop \cite{LeGallo:2016ga}. 
Yet, the structural dynamics during fast temperature quenching processes of overheated GSTs glasses and crystals is still unclear.\cite{sun:2006,Weinic:2007gq}. Indeed, when the heating-cooling cycle between two structural phases is closely observed, a variety of parameters (from quenching velocity to thermal dissipation and/or structural strain) dictate the out-of-equilibrium dynamical evolution in the energy phase space across either the amorphous-crystal or the crystalline-crystalline phase transformation\cite{Gawelda:2011iw}. Henceforth, the role of the quenching processes on the final structural configuration may not simply be a thermal dissipation, especially when the cooling rate are in the range of 10$^{12}$K/s or the heating stimulus is intense and ultrafast, i.e. in the ps time range.

To address these questions requires a description of the structural changes occurring at atomic level during the amorphous-crystalline phase transition or vice-versa. Obviously the amorphous character of one phase rules out the possibility of using direct long range order probe such as time-resolved X-ray diffraction. Unfortunately, this information entails to achieve unprecedented experimental and theoretical insights on the fundamental mechanisms at the base of the phase change transition related to the out-of-equilibrium dynamics of the local atomic structure during the quenching process. 

Scope of this work is to clarify the role of the ultrafast thermal strain dynamics in a CSL structure during the first instants of the heating-cooling cycle. Here by mean of first-principles multiple scattering simulations for interpreting time resolved X-ray absorption (tr-XAS) experiments, we unveil the microscopic structural and the dynamical changes occurring after the ultrafast heating of a nominal [GeTe(1nm)/Sb$_2$Te$_3$(3nm)]$_{15}$ CSL. 
The present results allow to unambiguously ascribe the distinct features of tr-XAS spectra to the dynamical structural strain occurring during the thermal quenching process. 

Recently, Ge L$_3$-edge XAS of GeTe based alloys\cite{Krbal:2011bh,Krbal:2012fv} have been interpreted using real-space \emph{ab-initio} multiple scattering simulations on crystalline and amorphous models. These studies have confirmed the effectiveness of the Ge L$_3$-edge as a spectroscopic probe to distinguish changes of the local atomic and electron charge distribution around the Ge photo-absorber.
\cite{Smolentsev:2007gq}.
Therefore, by extending the Ge L$_3$ absorption edge measurement to the time domain, unprecedented details about the local atomic structural dynamics during out-of-equilibrium states, like pre-melting phases and fast cooling processes, could be accessed. In addition, unlike X-ray diffraction, tr-XAS can be applied to both the crystalline and amorphous phases providing a unique information about the projected density of states (pDOS).

\section{Experiment}
In the present experiment we probe an as-grown [GeTe(1nm)/Sb$_2$Te$_3$(3nm)]$_{15}$CSL, which has been grown on the Sb-passivated surfaces of Si(111), ($\sqrt{3} \times \sqrt{3}$)R30$\degree$-Sb, at a substrate temperature of 230 $\degree$C, and capped with terminal layer for preventing oxidation. 
\cite{Casarin:2016dh,Momand:2015fz}.
The experiments have been carried out at the beamline BACH of the Elettra Synchrotron light source in Trieste, Italy, which operates an optical pump and X-ray probe technique capable of performing tr-XAS experiments with sub-nanosecond time resolution, hence making possible the direct observation of the structural evolution on ultrafast time scales. A general description of the setup is reported elsewhere.\cite{Stebel:2011bm}
In its standard multi-bunch operating mode, the Elettra storage ring delivers X-ray pulses with: (i) low intensity ($\sim$10$^3$ photons/pulse in a quasi-monochromatic beam), (ii) high repetition rate (500 MHz) and (iii) a $\sim$100 ps full-width-half-maximum (FWHM) photon pulse temporal profile.  This last parameter dictates the maximum time resolution of this experimental scheme.
A Ti-sapphire amplified laser source operating at $\sim$233 kHz repetition rate and synchronized with the storage ring radio frequency is used and delivers pump pulses of up to 50 fs pulses at 800 nm, with an energy/pulse of $\sim$6 $\mu$J. The time jitter of the laser with respect to the X-ray pulses is less then 5 ps, while the other relevant laser parameters are reported in table \ref{tab1}. The absorbed energy per pulse is calculated by measuring the sample transmittance response in the 0.1-1 eV energy range and extrapolating the response function value at 1.55 eV (see discussion in SI).

\begin{table}[b]
\begin{ruledtabular}
\begin{tabular}{c|rrrlrrrr}
Average power & 400  & mW &&     &  \\[1ex]
Wavelength ($\lambda$) & 800  & nm &&     &  \\[1ex]
Spot size & 250 & $\mu$m  &&   &  \\[1ex]
absorbed energy/pulse & 0.75 & $\mu$J && see SI      & \\
\end{tabular}
\end{ruledtabular}
\caption{\label{tab1}Laser parameters}
\end{table}

\begin{figure}[t!]
\captionsetup{justification=centerlast}
\centering
\includegraphics[width=0.95\columnwidth]{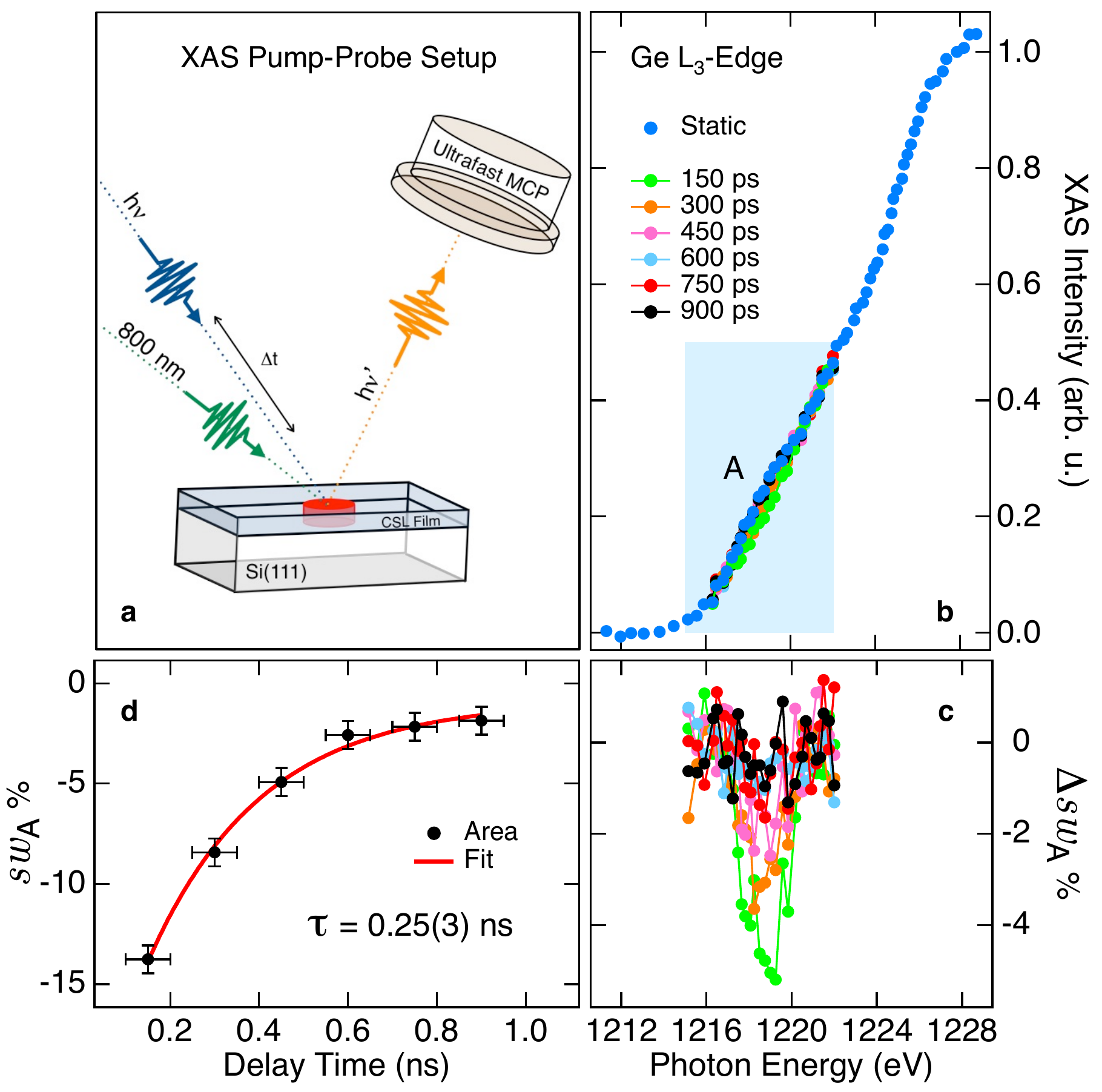}
\caption{(Color online) Panel (a) reports a simplified sketch of the setup: the 800 nm laser beam and the synchrotron X-ray pulses, synchronized with delay $\Delta$t, are both impinging on a CSL film grown on a Si substrate. Panel (b) shows a collection of Ge L$_3$ edges taken at different time delays with time step $\delta$t$\sim$150 ps, in comparison with a static Ge L$_3$ lineshape (dotted blue curve). Panel (c) shows a close-up of the shoulder A in the $\Delta$E=[1215-1220 eV] photon energy range (blue rectangle in panel (b)) plotted as $\Delta$sw$_A$(t) (see text). Panel (d): temporal dynamics of the spectral weight sw$_A$(t) (black dots, see text) fitted with a single exponential function.}
\label{fig1}
\end{figure}

A simplified sketch of the experimental pump-probe configuration is represented in panel (a) of Fig. \ref{fig1}.
The laser is focused on the sample by a 300 mm focal-length lens. Spatial overlap between X-ray and laser pulses is ensured by alignment of both pump and probe beams using a 100 $\mu$m pinhole \cite{Stebel:2011bm}.

The time resolved XAS Ge L$_3$-edge was probed in fluorescence mode using an Hamamatsu ultra-fast $\mu$-channel plate \cite{Stebel:2011bm} and acquired as a function of laser pulse time delay $\delta$t. 

The temporal overlapping of the pump and probe pulses defines the zero time delay ($\delta$t=0). 
The dynamics is measured in 150 ps time delay steps from t=-150 ps to t=900 ps for an overall time interval $\Delta$t of $\sim$1 ns.

\section{Results and discussion}
Fig. \ref{fig1}(b) reports a near edge region of Ge L$_3$-edge lineshape (blue dotted curve) measured across the 2p$_{3/2}$-sp absorption transition ($\sim$ 1220 eV) over a 15 eV photon energy range. The onset of the absorption edge is reported at equilibrium, i.e. without shining laser light on the sample, and after the pre-edge background removal and post-edge normalisation. 

A tiny spectral bump \emph{A} appears at the onset of the absorption threshold in the low energy region (blue box in \ref{fig1}(b)) being a direct signature of the specific local atomic Ge-Te coordination and electron charge distribution \cite{Krbal:2013vk}.

Upon laser illumination ($\delta t>0$), a sudden ($\ll$ 150 ps) but small change of the spectral weight of \emph{A} is observed.  

\begin{figure}[t!]
\captionsetup{justification=centerlast}
\centering
\includegraphics[width=0.98\columnwidth]{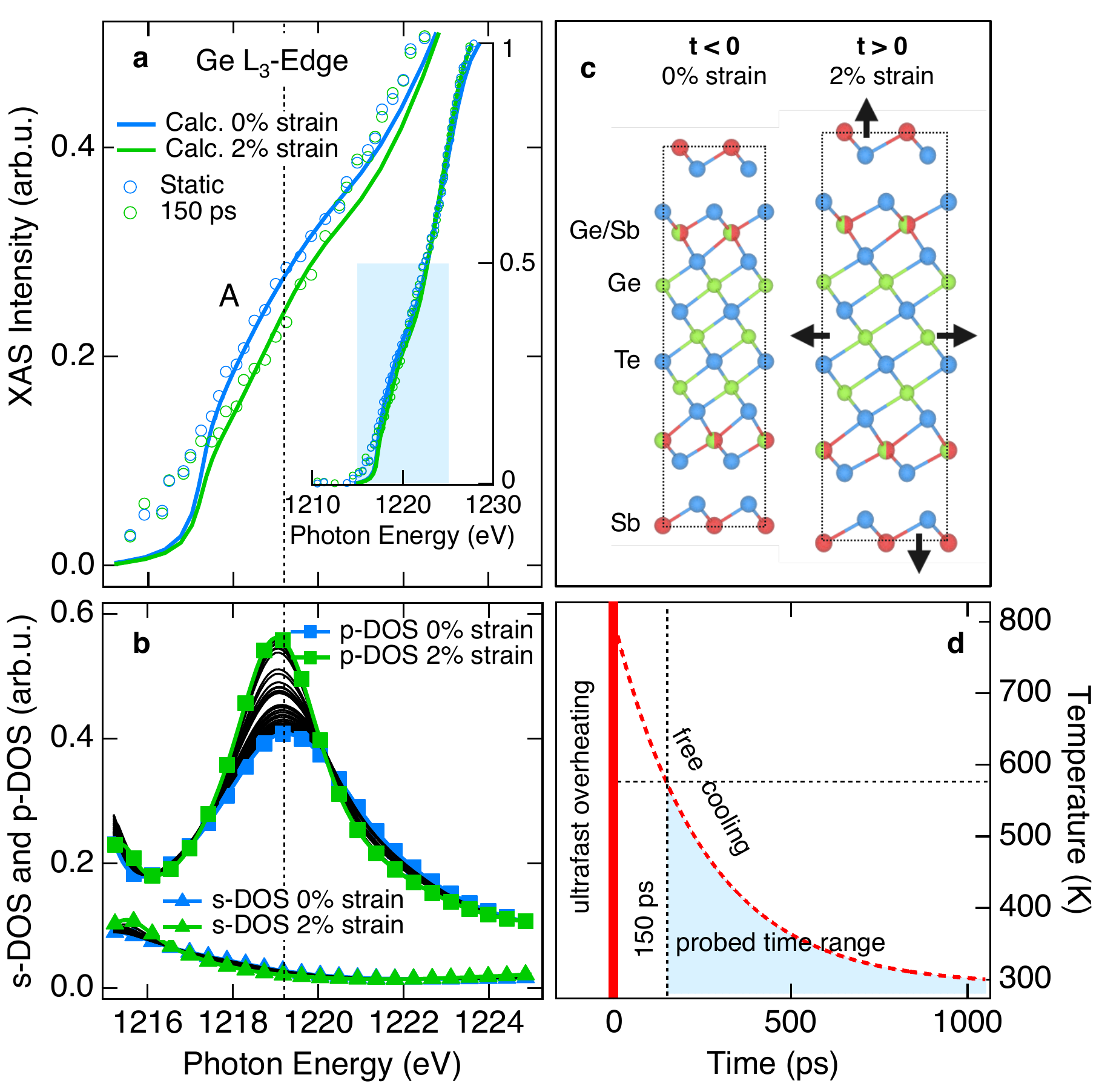}
\caption{(Color online) Panel (a) and inset therein: the experimental (open dots) Ge L$_3$ thresholds measured at time delays $\delta$t=0 (unpumped blue curves) and 150 ps (green dots), respectively, are compared to calculated (continuos curves) Ge L$_3$ thresholds. The simulated lineshapes have been calculated for no strain of 0 and 2\% strain of the lattice parameters, respectively. Panel (b) displays the Ge 3p- and 4s-DOS curves calculated for a series of strain [(0-2\%)] of the crystal lattice. Panel (c) sketches the CSL crystal cell before (equilibrium) and after (heated state) the arrival of the laser pulse. Therein a simplified version of the observed phenomenon is visually represented: the sudden change in temperature drives a lattice expansion of the CSL. The lattice will recover the initial state after complete thermal dissipation is achieved (T$_{in}$=T$_{fin}$). Panel (d) $\Delta$T calculated through eq. \ref{eq4}.}
\label{fig2}
\end{figure}

A collection of representative snapshots of the time evolution of the Ge-L$_3$ edge as a function of time delay after the laser excitation is shown superimposed to the equilibrium threshold. 

Magnified threshold changes over the bump energy region and time delays are displayed in Fig. \ref{fig1}(c)-(d), where the relative spectral weight change $\Delta sw_A(E)$ (Fig. \ref{fig1}c) and the integrated threshold change sw$_A$(t) (Fig. \ref{fig1}d) are reported. $\Delta sw_A(E)$ is calculated as the normalized difference between the threshold intensities as a fucntion of $\delta$t and of the reference static L$_3$ edge: 
\begin{equation}
\Delta sw_A(E)=\frac{sw_{t=0}(E)-sw_t(E)}{sw_{t=0}(E)}.
\label{eq1}
\end{equation}

sw$_A$(t) is then calculated by integrating $\Delta$sw$_A$(E) over the corresponding energy range: 

\begin{equation}
sw_A(t)=\int_{E1}^{E2}\Delta sw(E,\delta t)\Delta E. 
\label{eq2}
\end{equation}

It can be clearly seen that following the arrival of the laser pump pulse, $\Delta$sw$_A$(E) decreases and it reaches its minimum value after the first time delay step of $\sim$150 ps. In addition, considering the temporal evolution of sw$_A$(t), the initial decrease is followed by its almost complete recovery in about 1 ns. 

Since the overall time resolution is limited to the delay step that is comparable with the probe intrinsic time resolution ($\sim$ 100ps), spectral changes at shorter delays cannot be appreciated. 
Hence, the laser excitation and the resulting heating process are too fast to observe.

Ge-L$_3$ XAS corresponding to longer time delays confirms that the L$_3$ shoulder remains unchanged. This important observation indicates that the material completely recovered the initial state and the impulsive heating-fast cooling cycle is thus reversible. 

For the sake of clarity, in the following part of the discussion we will assume that the probed volume of the sample has an average uniform temperature depending $\delta$t. Thus temperature inhomogeneities, resulting from the depth dependent heat distribution due to the Beer-Lambert law of light absorption, are neglected. 
This assumption is validated by considering that the X-ray probe penetration depth at 1.2 KeV is $\sim$ 800 nm \cite{henke:online}. Thus the XAS is providing an averaged information about the local structure. 
In addition, since the thermal heat dissipation of CSL is relatively high, heat flow smears out the temperature distribution within the probed volume in few ps. It is worth to mention that the CSL/Si interface thermal resistance, which dictates the thermal flow from the CSL film through the bulk Si reservoir, is comparatively higher than the CSL thermal resistance.

The sw$_A$(t) temporal decay provides information on the dynamics, being sw$_A$(t) related to both the thermal dissipation and the structural relaxation. 
Accordingly, by using a one temperature model, the sw$_A$(t) was fitted with a single exponential function $a\exp^{-\frac{t}{\tau_0}}$, where $\tau_0$ is found to be $\sim$255 ps. The structure then relaxes with a cooling rate of 10$^{12}$ K/s, which is comparable to that expected for similar glass forming systems \cite{raoux2010phase,Jund:1997bz,Hegedus:2008cn}.

Even more interesting is to investigate the role of crystalline structure on the observed spectral changes. Thus, we have computed multiple scattering simulations of the Ge L$_3$ edge for a series of distorted structures of a known stable structure (displayed in Fig. \ref{fig2}(c) for positive and negative $\delta$t) via the ab-initio FEFF9 code.\cite{Rehr:2009eu,Rehr:2010tp} 

In Fig. \ref{fig2}(a) the calculated L$_3$ XAS for the undistorted structure and the largest distorted structure are displayed in the close-up energy region of the A bump and they are compared with the equilibrium and 150 ps delayed experimental data.
The inset of Fig. \ref{fig2}(a) displays the calculated and experimental L$_3$ spectra over an extended photon energy range.
The calculated spectra nicely reproduce the overall experimental line-shapes, while the maximum observable change in time of A is also well reproduced for a maximum 2\% lattice strain.

In Fig. \ref{fig2}(b), the calculated \emph{s-} and \emph{p-}symmetry DOSs of Ge are also reported. The \emph{p-}projected DOSs display a prominent feature peaked at 1219 eV (matching the photon energy range of the feature A in Fig. \ref{fig1}b), whose intensity changes as a function of lattice strain. Notably, a tiny  change of the \emph{s-}projected DOSs is observed, which is consistent with a small \emph{s-p} mixing.

This phenomenological analysis of the experimental data suggests that the detected changes measured at onset of the Ge L$_3$ absorption edge, between the sample at equilibrium and after the photoexcitations, i.e. feature A in Fig. \ref{fig1}b, should originates from lattice strains.  
This mechanisms is further supported by considering that the bonding overlap between the directional \emph{p} orbitals of Ge and the first nearest neighbours is strongly affected even by a small lattice expansion/contraction, while the almost spherical \emph{s} orbitals are only slightly perturbed.
On a side note, this observation is also relevant in terms of the ferroelectric properties of the medium, because stretched p-bonds can increase the local electric dipole contribution to the overall polarizability. 

Henceforth, consistently with the above scenario, the crystal structure 
undergoes a sudden lattice expansion corresponding to a fast temperature increase due to the absorption of the ultrafast pump pulse \cite{Thomsen:1986gf}. Then, both the out-of-equilibrium electronic/phonon subsystems and the lattice relax, following the heating thermal dissipation at a cooling rate of 10$^{12}$ K/s (Fig. \ref{fig2}(c-d)).

\begin{table}[b!]
\begin{ruledtabular}
\begin{tabular}{c|rrr|rrrr}
 & CSL (this work)&   && references&\\[1ex]
\hline\\
$\tau_0$ & 255  & ps &&       & \\[1ex]
$\alpha$$_L$ & 6 10$^{-5}$  & K$^{-1}$ && 2.44 10$^{-5}$ GST$_{225}$\cite{Park:2008dp,Kalb:2003vo}   &  \\[1ex]
$\delta \sigma$ & 2\%  & $\frac{\Delta L}{L}$ &&     &  \\[1ex]
$\Delta T$$_{/pulse}$ & $\sim$ 300 & K  &&   &  \\[1ex]
$\frac{\Delta T}{\tau_0}$ & $\sim$ 10$^{12}$  & K/s &&   ref. \cite[p.~261]{raoux2010phase}\cite{Jund:1997bz,Hegedus:2008cn}    & \\[1ex]
\end{tabular}
\end{ruledtabular}
\caption{\label{tab2} Thermoelastic parameters of the out-of-equilibrium CSL}
\end{table}

The average temperature distribution T(t) in the film is calculated from Fourier's law in one dimension, assuming that the electrons and phonons in the system remain in thermal equilibrium. 
This changing temperature distribution creates a structural strain that can be calculated as
\begin{equation}
\frac{\Delta L}{L}=\delta\sigma=-\alpha_L \delta T
\label{eq3}
\end{equation}
Since any assumption about the strength of the electron-electron and electron-phonon couplings is neglected, it is safe to assume thermal relaxation between the electron and phonon baths \cite{Eesley:1986gk}.
 
Considering a consistent thermodynamical approach, $\Delta$T can be calculated through 
\begin{equation}
\Delta Q=\rho V \int_{RT}^T c_p dT
\label{eq4}
\end{equation}
where $\rho$ is the film density \cite{Battaglia:2010bh}, $c_p$ the specific heat (see SI for a detailed discussion), V the heated volume, and $\Delta$Q the overall energy absorbed per laser pulse (table \ref{tab1}). This approach allows us to predict an average  temperature increase $\Delta T\cong$ 300 K, per pulse, of the heated volume. Consequently, $\alpha_L$ can be calculated for our CSL sample from equation \ref{eq4}, which results a factor $\sim$2.5 higher than the GST$_{225}$ case (both values are indicated in Table \ref{tab2}). This implies that, at least during the fast thermal quenching, the CSL structure is softer (higher $\alpha_L$) then the bulk case. 

Finally, it is important to underly that the extrapolated value of $\Delta$T, for time delays below 150 ps (Fig. \ref{fig2}(d)), does not exceed the melting temperature of the superlattice \cite{Yamada:1991eg}.
Combining this finding with the experimental observation that no phase transformation (e.g. melting) is observed by means of tr-XAS, we demonstrate that ultrafast optical overheating is a reversible process at this laser fluence regime and photon energy. 

\section{Conclusions}
In conclusion,  the dynamics of the Ge local atomic structure in a chalcogenide superlattice, during the thermal quenching phase of a reversible ultrafast heating - fast cooling cycle is revealed by time resolved XAS and first-principle theory modelling.

In addition to probing the atomic local structure, our approach reveals the significant impact of the lattice strain on the strength of bonds between atoms, from which strongly depends the quenching dynamics and, for example, the melting kinetics of a solid.

Our method is used here to (i) interpret the observed XAS spectral changes in terms of a dynamical microstructural picture of the Ge 4p-bonding relaxation and (ii) estimate relevant elastic properties of the out-of-equilibrium state of the CSL film.

Futhermore, by combining thermoelastic considerations and a microscopic multiple scattering approach we establish a direct connection between the structural microscopic evolution and the dielectric response in a CSL, which is fundamental for developing a microscopic theory for ultrafast phase transition and ultimately design new PCMs with improved performances. 

All together these results can open the route for future studies aimed to clarify the role of a transient structural strain on the strength of bonds between atoms in phase change materials in the proximity or even during a phase transition.

\section{ACKNOWLEDGMENTS}

M.M. acknowledges the support of the BACH beamline staff during the synchrotron experiments and Roberta Ciprian for insightful discussions. This work was supported by EU within FP7 project PASTRY [GA 317764].


\begin{thebibliography}{40}%
\makeatletter
\providecommand \@ifxundefined [1]{%
 \@ifx{#1\undefined}
}%
\providecommand \@ifnum [1]{%
 \ifnum #1\expandafter \@firstoftwo
 \else \expandafter \@secondoftwo
 \fi
}%
\providecommand \@ifx [1]{%
 \ifx #1\expandafter \@firstoftwo
 \else \expandafter \@secondoftwo
 \fi
}%
\providecommand \natexlab [1]{#1}%
\providecommand \enquote  [1]{``#1''}%
\providecommand \bibnamefont  [1]{#1}%
\providecommand \bibfnamefont [1]{#1}%
\providecommand \citenamefont [1]{#1}%
\providecommand \href@noop [0]{\@secondoftwo}%
\providecommand \href [0]{\begingroup \@sanitize@url \@href}%
\providecommand \@href[1]{\@@startlink{#1}\@@href}%
\providecommand \@@href[1]{\endgroup#1\@@endlink}%
\providecommand \@sanitize@url [0]{\catcode `\\12\catcode `\$12\catcode
  `\&12\catcode `\#12\catcode `\^12\catcode `\_12\catcode `\%12\relax}%
\providecommand \@@startlink[1]{}%
\providecommand \@@endlink[0]{}%
\providecommand \url  [0]{\begingroup\@sanitize@url \@url }%
\providecommand \@url [1]{\endgroup\@href {#1}{\urlprefix }}%
\providecommand \urlprefix  [0]{URL }%
\providecommand \Eprint [0]{\href }%
\providecommand \doibase [0]{http://dx.doi.org/}%
\providecommand \selectlanguage [0]{\@gobble}%
\providecommand \bibinfo  [0]{\@secondoftwo}%
\providecommand \bibfield  [0]{\@secondoftwo}%
\providecommand \translation [1]{[#1]}%
\providecommand \BibitemOpen [0]{}%
\providecommand \bibitemStop [0]{}%
\providecommand \bibitemNoStop [0]{.\EOS\space}%
\providecommand \EOS [0]{\spacefactor3000\relax}%
\providecommand \BibitemShut  [1]{\csname bibitem#1\endcsname}%
\let\auto@bib@innerbib\@empty
\bibitem [{\citenamefont {Wuttig}\ and\ \citenamefont
  {Yamada}(2007)}]{Wuttig:2007bs}%
  \BibitemOpen
  \bibfield  {author} {\bibinfo {author} {\bibfnamefont {M.}~\bibnamefont
  {Wuttig}}\ and\ \bibinfo {author} {\bibfnamefont {N.}~\bibnamefont
  {Yamada}},\ }\href {\doibase 10.1038/nmat2009} {\bibfield  {journal}
  {\bibinfo  {journal} {Nature Materials}\ }\textbf {\bibinfo {volume} {6}},\
  \bibinfo {pages} {824} (\bibinfo {year} {2007})}\BibitemShut {NoStop}%
\bibitem [{\citenamefont {Lankhorst}\ \emph {et~al.}(2005)\citenamefont
  {Lankhorst}, \citenamefont {Ketelaars},\ and\ \citenamefont
  {Wolters}}]{Lankhorst:2005cg}%
  \BibitemOpen
  \bibfield  {author} {\bibinfo {author} {\bibfnamefont {M.~H.~R.}\
  \bibnamefont {Lankhorst}}, \bibinfo {author} {\bibfnamefont {B.~W. S. M.~M.}\
  \bibnamefont {Ketelaars}}, \ and\ \bibinfo {author} {\bibfnamefont
  {R.~A.~M.}\ \bibnamefont {Wolters}},\ }\href {\doibase 10.1038/nmat1350}
  {\bibfield  {journal} {\bibinfo  {journal} {Nature Materials}\ }\textbf
  {\bibinfo {volume} {4}},\ \bibinfo {pages} {347} (\bibinfo {year}
  {2005})}\BibitemShut {NoStop}%
\bibitem [{\citenamefont {Tominaga}\ \emph {et~al.}(2008)\citenamefont
  {Tominaga}, \citenamefont {Fons}, \citenamefont {Kolobov}, \citenamefont
  {Shima}, \citenamefont {Chong}, \citenamefont {Zhao}, \citenamefont {Lee},\
  and\ \citenamefont {Shi}}]{Tominaga:2008fh}%
  \BibitemOpen
  \bibfield  {author} {\bibinfo {author} {\bibfnamefont {J.}~\bibnamefont
  {Tominaga}}, \bibinfo {author} {\bibfnamefont {P.}~\bibnamefont {Fons}},
  \bibinfo {author} {\bibfnamefont {A.}~\bibnamefont {Kolobov}}, \bibinfo
  {author} {\bibfnamefont {T.}~\bibnamefont {Shima}}, \bibinfo {author}
  {\bibfnamefont {T.~C.}\ \bibnamefont {Chong}}, \bibinfo {author}
  {\bibfnamefont {R.}~\bibnamefont {Zhao}}, \bibinfo {author} {\bibfnamefont
  {H.~K.}\ \bibnamefont {Lee}}, \ and\ \bibinfo {author} {\bibfnamefont
  {L.}~\bibnamefont {Shi}},\ }\href {\doibase 10.1143/JJAP.47.5763} {\bibfield
  {journal} {\bibinfo  {journal} {Japanese Journal of Applied Physics}\
  }\textbf {\bibinfo {volume} {47}},\ \bibinfo {pages} {5763} (\bibinfo {year}
  {2008})}\BibitemShut {NoStop}%
\bibitem [{\citenamefont {Simpson}\ \emph {et~al.}(2011)\citenamefont
  {Simpson}, \citenamefont {Fons}, \citenamefont {Kolobov}, \citenamefont
  {Fukaya}, \citenamefont {Krbal}, \citenamefont {Yagi},\ and\ \citenamefont
  {Tominaga}}]{Simpson:2011kh}%
  \BibitemOpen
  \bibfield  {author} {\bibinfo {author} {\bibfnamefont {R.}~\bibnamefont
  {Simpson}}, \bibinfo {author} {\bibfnamefont {P.}~\bibnamefont {Fons}},
  \bibinfo {author} {\bibfnamefont {A.}~\bibnamefont {Kolobov}}, \bibinfo
  {author} {\bibfnamefont {T.}~\bibnamefont {Fukaya}}, \bibinfo {author}
  {\bibfnamefont {M.}~\bibnamefont {Krbal}}, \bibinfo {author} {\bibfnamefont
  {T.}~\bibnamefont {Yagi}}, \ and\ \bibinfo {author} {\bibfnamefont
  {J.}~\bibnamefont {Tominaga}},\ }\href {\doibase 10.1038/nnano.2011.96}
  {\bibfield  {journal} {\bibinfo  {journal} {Nature Nanotechnology}\ }\textbf
  {\bibinfo {volume} {6}},\ \bibinfo {pages} {501} (\bibinfo {year}
  {2011})}\BibitemShut {NoStop}%
\bibitem [{\citenamefont {Baker}\ \emph {et~al.}(2006)\citenamefont {Baker},
  \citenamefont {Paesler}, \citenamefont {Lucovsky}, \citenamefont {Agarwal},\
  and\ \citenamefont {{Taylor, P.C.}}}]{Baker:2006br}%
  \BibitemOpen
  \bibfield  {author} {\bibinfo {author} {\bibfnamefont {D.~A.}\ \bibnamefont
  {Baker}}, \bibinfo {author} {\bibfnamefont {M.~A.}\ \bibnamefont {Paesler}},
  \bibinfo {author} {\bibfnamefont {G.}~\bibnamefont {Lucovsky}}, \bibinfo
  {author} {\bibfnamefont {S.~C.}\ \bibnamefont {Agarwal}}, \ and\ \bibinfo
  {author} {\bibnamefont {{Taylor, P.C.}}},\ }\href {\doibase
  10.1103/PhysRevLett.96.255501} {\bibfield  {journal} {\bibinfo  {journal}
  {Physical Review Letters}\ }\textbf {\bibinfo {volume} {96}},\ \bibinfo
  {pages} {255501} (\bibinfo {year} {2006})}\BibitemShut {NoStop}%
\bibitem [{\citenamefont {Yu}\ and\ \citenamefont
  {Robertson}(2015)}]{Yu:2015jw}%
  \BibitemOpen
  \bibfield  {author} {\bibinfo {author} {\bibfnamefont {X.}~\bibnamefont
  {Yu}}\ and\ \bibinfo {author} {\bibfnamefont {J.}~\bibnamefont {Robertson}},\
  }\href {\doibase 10.1038/srep12612} {\bibfield  {journal} {\bibinfo
  {journal} {Scientific Reports}\ }\textbf {\bibinfo {volume} {5}},\ \bibinfo
  {pages} {12612} (\bibinfo {year} {2015})}\BibitemShut {NoStop}%
\bibitem [{\citenamefont {Sun}\ \emph {et~al.}(2006)\citenamefont {Sun},
  \citenamefont {Zhou},\ and\ \citenamefont {Ahuja}}]{Sun:2006eb}%
  \BibitemOpen
  \bibfield  {author} {\bibinfo {author} {\bibfnamefont {Z.}~\bibnamefont
  {Sun}}, \bibinfo {author} {\bibfnamefont {J.}~\bibnamefont {Zhou}}, \ and\
  \bibinfo {author} {\bibfnamefont {R.}~\bibnamefont {Ahuja}},\ }\href
  {http://link.aps.org/doi/10.1103/PhysRevLett.96.055507} {\bibfield  {journal}
  {\bibinfo  {journal} {Physical Review Letters}\ }\textbf {\bibinfo {volume}
  {96}},\ \bibinfo {pages} {055507} (\bibinfo {year} {2006})}\BibitemShut
  {NoStop}%
\bibitem [{\citenamefont {We{\l}nic}\ \emph {et~al.}(2007)\citenamefont
  {We{\l}nic}, \citenamefont {Botti}, \citenamefont {Reining},\ and\
  \citenamefont {Wuttig}}]{Weinic:2007gq}%
  \BibitemOpen
  \bibfield  {author} {\bibinfo {author} {\bibfnamefont {W.}~\bibnamefont
  {We{\l}nic}}, \bibinfo {author} {\bibfnamefont {S.}~\bibnamefont {Botti}},
  \bibinfo {author} {\bibfnamefont {L.}~\bibnamefont {Reining}}, \ and\
  \bibinfo {author} {\bibfnamefont {M.}~\bibnamefont {Wuttig}},\ }\href
  {http://link.aps.org/doi/10.1103/PhysRevLett.98.236403} {\bibfield  {journal}
  {\bibinfo  {journal} {Physical Review Letters}\ }\textbf {\bibinfo {volume}
  {98}},\ \bibinfo {pages} {236403} (\bibinfo {year} {2007})}\BibitemShut
  {NoStop}%
\bibitem [{\citenamefont {Sun}\ \emph {et~al.}(2007)\citenamefont {Sun},
  \citenamefont {Zhou},\ and\ \citenamefont {Ahuja}}]{sun:2006}%
  \BibitemOpen
  \bibfield  {author} {\bibinfo {author} {\bibfnamefont {Z.}~\bibnamefont
  {Sun}}, \bibinfo {author} {\bibfnamefont {J.}~\bibnamefont {Zhou}}, \ and\
  \bibinfo {author} {\bibfnamefont {R.}~\bibnamefont {Ahuja}},\ }\href
  {\doibase 10.1103/PhysRevLett.98.055505} {\bibfield  {journal} {\bibinfo
  {journal} {Physical Review Letters}\ }\textbf {\bibinfo {volume} {98}},\
  \bibinfo {pages} {055505} (\bibinfo {year} {2007})}\BibitemShut {NoStop}%
\bibitem [{\citenamefont {Klein}\ \emph {et~al.}(2008)\citenamefont {Klein},
  \citenamefont {Dieker}, \citenamefont {Sp{\"a}th}, \citenamefont {Fons},\
  and\ \citenamefont {Kolobov}}]{Klein:2008kk}%
  \BibitemOpen
  \bibfield  {author} {\bibinfo {author} {\bibfnamefont {A.}~\bibnamefont
  {Klein}}, \bibinfo {author} {\bibfnamefont {H.}~\bibnamefont {Dieker}},
  \bibinfo {author} {\bibfnamefont {B.}~\bibnamefont {Sp{\"a}th}}, \bibinfo
  {author} {\bibfnamefont {P.}~\bibnamefont {Fons}}, \ and\ \bibinfo {author}
  {\bibfnamefont {A.}~\bibnamefont {Kolobov}},\ }\href {\doibase
  10.1103/PhysRevLett.100.016402} {\bibfield  {journal} {\bibinfo  {journal}
  {Physical Review}\ } (\bibinfo {year} {2008}),\
  10.1103/PhysRevLett.100.016402}\BibitemShut {NoStop}%
\bibitem [{\citenamefont {Caravati}\ \emph {et~al.}(2009)\citenamefont
  {Caravati}, \citenamefont {Bernasconi}, \citenamefont {K{\"u}hne},
  \citenamefont {Krack},\ and\ \citenamefont {Parrinello}}]{Caravati:2009hp}%
  \BibitemOpen
  \bibfield  {author} {\bibinfo {author} {\bibfnamefont {S.}~\bibnamefont
  {Caravati}}, \bibinfo {author} {\bibfnamefont {M.}~\bibnamefont
  {Bernasconi}}, \bibinfo {author} {\bibfnamefont {T.~D.}\ \bibnamefont
  {K{\"u}hne}}, \bibinfo {author} {\bibfnamefont {M.}~\bibnamefont {Krack}}, \
  and\ \bibinfo {author} {\bibfnamefont {M.}~\bibnamefont {Parrinello}},\
  }\href {\doibase 10.1103/PhysRevLett.102.205502} {\bibfield  {journal}
  {\bibinfo  {journal} {Physical Review Letters}\ }\textbf {\bibinfo {volume}
  {102}},\ \bibinfo {pages} {205502} (\bibinfo {year} {2009})}\BibitemShut
  {NoStop}%
\bibitem [{\citenamefont {Lee}\ and\ \citenamefont
  {Henisch}(1972)}]{Lee:1972gx}%
  \BibitemOpen
  \bibfield  {author} {\bibinfo {author} {\bibfnamefont {S.~H.}\ \bibnamefont
  {Lee}}\ and\ \bibinfo {author} {\bibfnamefont {H.~K.}\ \bibnamefont
  {Henisch}},\ }\href {\doibase 10.1016/0022-3093(72)90002-6} {\bibfield
  {journal} {\bibinfo  {journal} {Journal of Non-Crystalline Solids}\ }\textbf
  {\bibinfo {volume} {11}},\ \bibinfo {pages} {192} (\bibinfo {year}
  {1972})}\BibitemShut {NoStop}%
\bibitem [{\citenamefont {B{\"o}er}\ and\ \citenamefont
  {Ovshinsky}(1970)}]{Boer:1970hj}%
  \BibitemOpen
  \bibfield  {author} {\bibinfo {author} {\bibfnamefont {K.~W.}\ \bibnamefont
  {B{\"o}er}}\ and\ \bibinfo {author} {\bibfnamefont {S.~R.}\ \bibnamefont
  {Ovshinsky}},\ }\href {\doibase 10.1063/1.1659281} {\bibfield  {journal}
  {\bibinfo  {journal} {Journal of Applied Physics}\ }\textbf {\bibinfo
  {volume} {41}},\ \bibinfo {pages} {2675} (\bibinfo {year}
  {1970})}\BibitemShut {NoStop}%
\bibitem [{\citenamefont {Makino}\ \emph {et~al.}(2012)\citenamefont {Makino},
  \citenamefont {Tominaga}, \citenamefont {Kolobov}, \citenamefont {Fons},\
  and\ \citenamefont {Hase}}]{Makino:2012bq}%
  \BibitemOpen
  \bibfield  {author} {\bibinfo {author} {\bibfnamefont {K.}~\bibnamefont
  {Makino}}, \bibinfo {author} {\bibfnamefont {J.}~\bibnamefont {Tominaga}},
  \bibinfo {author} {\bibfnamefont {A.~V.}\ \bibnamefont {Kolobov}}, \bibinfo
  {author} {\bibfnamefont {P.}~\bibnamefont {Fons}}, \ and\ \bibinfo {author}
  {\bibfnamefont {M.}~\bibnamefont {Hase}},\ }\href {\doibase
  10.1063/1.4768785} {\bibfield  {journal} {\bibinfo  {journal} {Applied
  Physics Letters}\ }\textbf {\bibinfo {volume} {101}},\ \bibinfo {pages}
  {232101} (\bibinfo {year} {2012})}\BibitemShut {NoStop}%
\bibitem [{\citenamefont {Le~Gallo}\ \emph {et~al.}(2016)\citenamefont
  {Le~Gallo}, \citenamefont {Athmanathan}, \citenamefont {Krebs},\ and\
  \citenamefont {Sebastian}}]{LeGallo:2016ga}%
  \BibitemOpen
  \bibfield  {author} {\bibinfo {author} {\bibfnamefont {M.}~\bibnamefont
  {Le~Gallo}}, \bibinfo {author} {\bibfnamefont {A.}~\bibnamefont
  {Athmanathan}}, \bibinfo {author} {\bibfnamefont {D.}~\bibnamefont {Krebs}},
  \ and\ \bibinfo {author} {\bibfnamefont {A.}~\bibnamefont {Sebastian}},\
  }\href {\doibase 10.1063/1.4938532} {\bibfield  {journal} {\bibinfo
  {journal} {Journal of Applied Physics}\ }\textbf {\bibinfo {volume} {119}},\
  \bibinfo {pages} {025704} (\bibinfo {year} {2016})}\BibitemShut {NoStop}%
\bibitem [{\citenamefont {Bruns}\ \emph {et~al.}(2009)\citenamefont {Bruns},
  \citenamefont {Merkelbach}, \citenamefont {Schlockermann}, \citenamefont
  {Salinga}, \citenamefont {Wuttig}, \citenamefont {Happ}, \citenamefont
  {Philipp},\ and\ \citenamefont {Kund}}]{Bruns:2009gj}%
  \BibitemOpen
  \bibfield  {author} {\bibinfo {author} {\bibfnamefont {G.}~\bibnamefont
  {Bruns}}, \bibinfo {author} {\bibfnamefont {P.}~\bibnamefont {Merkelbach}},
  \bibinfo {author} {\bibfnamefont {C.}~\bibnamefont {Schlockermann}}, \bibinfo
  {author} {\bibfnamefont {M.}~\bibnamefont {Salinga}}, \bibinfo {author}
  {\bibfnamefont {M.}~\bibnamefont {Wuttig}}, \bibinfo {author} {\bibfnamefont
  {T.~D.}\ \bibnamefont {Happ}}, \bibinfo {author} {\bibfnamefont {J.~B.}\
  \bibnamefont {Philipp}}, \ and\ \bibinfo {author} {\bibfnamefont
  {M.}~\bibnamefont {Kund}},\ }\href {\doibase 10.1063/1.3191670} {\bibfield
  {journal} {\bibinfo  {journal} {Applied Physics Letters}\ }\textbf {\bibinfo
  {volume} {95}},\ \bibinfo {pages} {043108} (\bibinfo {year}
  {2009})}\BibitemShut {NoStop}%
\bibitem [{\citenamefont {Krebs}\ \emph {et~al.}(2009)\citenamefont {Krebs},
  \citenamefont {Raoux}, \citenamefont {Rettner}, \citenamefont {Burr},
  \citenamefont {Salinga},\ and\ \citenamefont {Wuttig}}]{Krebs:2009is}%
  \BibitemOpen
  \bibfield  {author} {\bibinfo {author} {\bibfnamefont {D.}~\bibnamefont
  {Krebs}}, \bibinfo {author} {\bibfnamefont {S.}~\bibnamefont {Raoux}},
  \bibinfo {author} {\bibfnamefont {C.~T.}\ \bibnamefont {Rettner}}, \bibinfo
  {author} {\bibfnamefont {G.~W.}\ \bibnamefont {Burr}}, \bibinfo {author}
  {\bibfnamefont {M.}~\bibnamefont {Salinga}}, \ and\ \bibinfo {author}
  {\bibfnamefont {M.}~\bibnamefont {Wuttig}},\ }\href {\doibase
  10.1063/1.3210792} {\bibfield  {journal} {\bibinfo  {journal} {Applied
  Physics Letters}\ }\textbf {\bibinfo {volume} {95}},\ \bibinfo {pages}
  {082101} (\bibinfo {year} {2009})}\BibitemShut {NoStop}%
\bibitem [{\citenamefont {Kohary}\ and\ \citenamefont
  {Wright}(2011)}]{Kohary:2011bn}%
  \BibitemOpen
  \bibfield  {author} {\bibinfo {author} {\bibfnamefont {K.}~\bibnamefont
  {Kohary}}\ and\ \bibinfo {author} {\bibfnamefont {C.~D.}\ \bibnamefont
  {Wright}},\ }\href {\doibase 10.1063/1.3595408} {\bibfield  {journal}
  {\bibinfo  {journal} {Applied Physics Letters}\ }\textbf {\bibinfo {volume}
  {98}},\ \bibinfo {pages} {223102} (\bibinfo {year} {2011})}\BibitemShut
  {NoStop}%
\bibitem [{\citenamefont {V{\'a}zquez~Diosdado}\ \emph
  {et~al.}(2012)\citenamefont {V{\'a}zquez~Diosdado}, \citenamefont {Ashwin},
  \citenamefont {Kohary},\ and\ \citenamefont
  {Wright}}]{VazquezDiosdado:2012cv}%
  \BibitemOpen
  \bibfield  {author} {\bibinfo {author} {\bibfnamefont {J.~A.}\ \bibnamefont
  {V{\'a}zquez~Diosdado}}, \bibinfo {author} {\bibfnamefont {P.}~\bibnamefont
  {Ashwin}}, \bibinfo {author} {\bibfnamefont {K.~I.}\ \bibnamefont {Kohary}},
  \ and\ \bibinfo {author} {\bibfnamefont {C.~D.}\ \bibnamefont {Wright}},\
  }\href {\doibase 10.1063/1.4729551} {\bibfield  {journal} {\bibinfo
  {journal} {Applied Physics Letters}\ }\textbf {\bibinfo {volume} {100}},\
  \bibinfo {pages} {253105} (\bibinfo {year} {2012})}\BibitemShut {NoStop}%
\bibitem [{\citenamefont {Cao}\ \emph {et~al.}(2015)\citenamefont {Cao},
  \citenamefont {Wu}, \citenamefont {Zhu}, \citenamefont {Ji}, \citenamefont
  {Zheng}, \citenamefont {Song}, \citenamefont {Rao}, \citenamefont {Song},
  \citenamefont {Ma},\ and\ \citenamefont {Xu}}]{Cao:2015bq}%
  \BibitemOpen
  \bibfield  {author} {\bibinfo {author} {\bibfnamefont {L.}~\bibnamefont
  {Cao}}, \bibinfo {author} {\bibfnamefont {L.}~\bibnamefont {Wu}}, \bibinfo
  {author} {\bibfnamefont {W.}~\bibnamefont {Zhu}}, \bibinfo {author}
  {\bibfnamefont {X.}~\bibnamefont {Ji}}, \bibinfo {author} {\bibfnamefont
  {Y.}~\bibnamefont {Zheng}}, \bibinfo {author} {\bibfnamefont
  {Z.}~\bibnamefont {Song}}, \bibinfo {author} {\bibfnamefont {F.}~\bibnamefont
  {Rao}}, \bibinfo {author} {\bibfnamefont {S.}~\bibnamefont {Song}}, \bibinfo
  {author} {\bibfnamefont {Z.}~\bibnamefont {Ma}}, \ and\ \bibinfo {author}
  {\bibfnamefont {L.}~\bibnamefont {Xu}},\ }\href {\doibase 10.1063/1.4937603}
  {\bibfield  {journal} {\bibinfo  {journal} {Applied Physics Letters}\
  }\textbf {\bibinfo {volume} {107}},\ \bibinfo {pages} {242101} (\bibinfo
  {year} {2015})}\BibitemShut {NoStop}%
\bibitem [{\citenamefont {Gawelda}\ \emph {et~al.}(2011)\citenamefont
  {Gawelda}, \citenamefont {Siegel}, \citenamefont {Afonso}, \citenamefont
  {Plausinaitiene}, \citenamefont {Abrutis},\ and\ \citenamefont
  {Wiemer}}]{Gawelda:2011iw}%
  \BibitemOpen
  \bibfield  {author} {\bibinfo {author} {\bibfnamefont {W.}~\bibnamefont
  {Gawelda}}, \bibinfo {author} {\bibfnamefont {J.}~\bibnamefont {Siegel}},
  \bibinfo {author} {\bibfnamefont {C.~N.}\ \bibnamefont {Afonso}}, \bibinfo
  {author} {\bibfnamefont {V.}~\bibnamefont {Plausinaitiene}}, \bibinfo
  {author} {\bibfnamefont {A.}~\bibnamefont {Abrutis}}, \ and\ \bibinfo
  {author} {\bibfnamefont {C.}~\bibnamefont {Wiemer}},\ }\href {\doibase
  10.1063/1.3596562} {\bibfield  {journal} {\bibinfo  {journal} {Journal of
  Applied Physics}\ }\textbf {\bibinfo {volume} {109}},\ \bibinfo {pages}
  {123102} (\bibinfo {year} {2011})}\BibitemShut {NoStop}%
\bibitem [{\citenamefont {Krbal}\ \emph {et~al.}(2011)\citenamefont {Krbal},
  \citenamefont {Kolobov}, \citenamefont {Fons}, \citenamefont {Tominaga},
  \citenamefont {Elliott}, \citenamefont {Heged{\"u}s},\ and\ \citenamefont
  {Uruga}}]{Krbal:2011bh}%
  \BibitemOpen
  \bibfield  {author} {\bibinfo {author} {\bibfnamefont {M.}~\bibnamefont
  {Krbal}}, \bibinfo {author} {\bibfnamefont {A.}~\bibnamefont {Kolobov}},
  \bibinfo {author} {\bibfnamefont {P.}~\bibnamefont {Fons}}, \bibinfo {author}
  {\bibfnamefont {J.}~\bibnamefont {Tominaga}}, \bibinfo {author}
  {\bibfnamefont {S.~R.}\ \bibnamefont {Elliott}}, \bibinfo {author}
  {\bibfnamefont {J.}~\bibnamefont {Heged{\"u}s}}, \ and\ \bibinfo {author}
  {\bibfnamefont {T.}~\bibnamefont {Uruga}},\ }\href {\doibase
  10.1103/PhysRevB.83.054203} {\bibfield  {journal} {\bibinfo  {journal}
  {Physical Review B - Condensed Matter and Materials Physics}\ }\textbf
  {\bibinfo {volume} {83}},\ \bibinfo {pages} {054203} (\bibinfo {year}
  {2011})}\BibitemShut {NoStop}%
\bibitem [{\citenamefont {Krbal}\ \emph {et~al.}(2012)\citenamefont {Krbal},
  \citenamefont {Kolobov}, \citenamefont {Fons}, \citenamefont {Tominaga},
  \citenamefont {Elliott}, \citenamefont {Heged{\"u}s}, \citenamefont
  {Guissani}, \citenamefont {Perumal}, \citenamefont {Calarco}, \citenamefont
  {Matsunaga}, \citenamefont {Yamada}, \citenamefont {Nitta},\ and\
  \citenamefont {Uruga}}]{Krbal:2012fv}%
  \BibitemOpen
  \bibfield  {author} {\bibinfo {author} {\bibfnamefont {M.}~\bibnamefont
  {Krbal}}, \bibinfo {author} {\bibfnamefont {A.}~\bibnamefont {Kolobov}},
  \bibinfo {author} {\bibfnamefont {P.}~\bibnamefont {Fons}}, \bibinfo {author}
  {\bibfnamefont {J.}~\bibnamefont {Tominaga}}, \bibinfo {author}
  {\bibfnamefont {S.~R.}\ \bibnamefont {Elliott}}, \bibinfo {author}
  {\bibfnamefont {J.}~\bibnamefont {Heged{\"u}s}}, \bibinfo {author}
  {\bibfnamefont {A.}~\bibnamefont {Guissani}}, \bibinfo {author}
  {\bibfnamefont {K.}~\bibnamefont {Perumal}}, \bibinfo {author} {\bibfnamefont
  {R.}~\bibnamefont {Calarco}}, \bibinfo {author} {\bibfnamefont
  {T.}~\bibnamefont {Matsunaga}}, \bibinfo {author} {\bibfnamefont
  {N.}~\bibnamefont {Yamada}}, \bibinfo {author} {\bibfnamefont
  {K.}~\bibnamefont {Nitta}}, \ and\ \bibinfo {author} {\bibfnamefont
  {T.}~\bibnamefont {Uruga}},\ }\href {\doibase 10.1103/PhysRevB.86.045212}
  {\bibfield  {journal} {\bibinfo  {journal} {Physical Review B - Condensed
  Matter and Materials Physics}\ }\textbf {\bibinfo {volume} {86}},\ \bibinfo
  {pages} {045212} (\bibinfo {year} {2012})}\BibitemShut {NoStop}%
\bibitem [{\citenamefont {Smolentsev}\ \emph {et~al.}(2007)\citenamefont
  {Smolentsev}, \citenamefont {Soldatov},\ and\ \citenamefont
  {Feiters}}]{Smolentsev:2007gq}%
  \BibitemOpen
  \bibfield  {author} {\bibinfo {author} {\bibfnamefont {G.}~\bibnamefont
  {Smolentsev}}, \bibinfo {author} {\bibfnamefont {A.~V.}\ \bibnamefont
  {Soldatov}}, \ and\ \bibinfo {author} {\bibfnamefont {M.~C.}\ \bibnamefont
  {Feiters}},\ }\href {http://link.aps.org/doi/10.1103/PhysRevB.75.144106}
  {\bibfield  {journal} {\bibinfo  {journal} {Physical Review B}\ }\textbf
  {\bibinfo {volume} {75}},\ \bibinfo {pages} {144106} (\bibinfo {year}
  {2007})}\BibitemShut {NoStop}%
\bibitem [{\citenamefont {Casarin}\ \emph {et~al.}(2016)\citenamefont
  {Casarin}, \citenamefont {Caretta}, \citenamefont {Momand}, \citenamefont
  {Kooi}, \citenamefont {Verheijen}, \citenamefont {Bragaglia}, \citenamefont
  {Calarco}, \citenamefont {Chukalina}, \citenamefont {Yu}, \citenamefont
  {Robertson}, \citenamefont {Lange}, \citenamefont {Wuttig}, \citenamefont
  {Redaelli}, \citenamefont {Varesi}, \citenamefont {Parmigiani},\ and\
  \citenamefont {Malvestuto}}]{Casarin:2016dh}%
  \BibitemOpen
  \bibfield  {author} {\bibinfo {author} {\bibfnamefont {B.}~\bibnamefont
  {Casarin}}, \bibinfo {author} {\bibfnamefont {A.}~\bibnamefont {Caretta}},
  \bibinfo {author} {\bibfnamefont {J.}~\bibnamefont {Momand}}, \bibinfo
  {author} {\bibfnamefont {B.~J.}\ \bibnamefont {Kooi}}, \bibinfo {author}
  {\bibfnamefont {M.~A.}\ \bibnamefont {Verheijen}}, \bibinfo {author}
  {\bibfnamefont {V.}~\bibnamefont {Bragaglia}}, \bibinfo {author}
  {\bibfnamefont {R.}~\bibnamefont {Calarco}}, \bibinfo {author} {\bibfnamefont
  {M.}~\bibnamefont {Chukalina}}, \bibinfo {author} {\bibfnamefont
  {X.}~\bibnamefont {Yu}}, \bibinfo {author} {\bibfnamefont {J.}~\bibnamefont
  {Robertson}}, \bibinfo {author} {\bibfnamefont {F.~R.~L.}\ \bibnamefont
  {Lange}}, \bibinfo {author} {\bibfnamefont {M.}~\bibnamefont {Wuttig}},
  \bibinfo {author} {\bibfnamefont {A.}~\bibnamefont {Redaelli}}, \bibinfo
  {author} {\bibfnamefont {E.}~\bibnamefont {Varesi}}, \bibinfo {author}
  {\bibfnamefont {F.}~\bibnamefont {Parmigiani}}, \ and\ \bibinfo {author}
  {\bibfnamefont {M.}~\bibnamefont {Malvestuto}},\ }\href {\doibase
  10.1038/srep22353} {\bibfield  {journal} {\bibinfo  {journal} {Scientific
  Reports}\ }\textbf {\bibinfo {volume} {6}},\ \bibinfo {pages} {22353}
  (\bibinfo {year} {2016})}\BibitemShut {NoStop}%
\bibitem [{\citenamefont {Momand}\ \emph {et~al.}(2015)\citenamefont {Momand},
  \citenamefont {Wang}, \citenamefont {Boschker}, \citenamefont {Verheijen},
  \citenamefont {Calarco},\ and\ \citenamefont {Kooi}}]{Momand:2015fz}%
  \BibitemOpen
  \bibfield  {author} {\bibinfo {author} {\bibfnamefont {J.}~\bibnamefont
  {Momand}}, \bibinfo {author} {\bibfnamefont {R.}~\bibnamefont {Wang}},
  \bibinfo {author} {\bibfnamefont {J.}~\bibnamefont {Boschker}}, \bibinfo
  {author} {\bibfnamefont {M.~A.}\ \bibnamefont {Verheijen}}, \bibinfo {author}
  {\bibfnamefont {R.}~\bibnamefont {Calarco}}, \ and\ \bibinfo {author}
  {\bibfnamefont {B.~J.}\ \bibnamefont {Kooi}},\ }\href {\doibase
  10.1039/C5NR04530D} {\bibfield  {journal} {\bibinfo  {journal} {Nanoscale}\
  }\textbf {\bibinfo {volume} {7}},\ \bibinfo {pages} {19136} (\bibinfo {year}
  {2015})}\BibitemShut {NoStop}%
\bibitem [{\citenamefont {Stebel}\ \emph {et~al.}(2011)\citenamefont {Stebel},
  \citenamefont {Malvestuto}, \citenamefont {Capogrosso}, \citenamefont
  {Sigalotti}, \citenamefont {Ressel}, \citenamefont {Bondino}, \citenamefont
  {Magnano}, \citenamefont {Cautero},\ and\ \citenamefont
  {Parmigiani}}]{Stebel:2011bm}%
  \BibitemOpen
  \bibfield  {author} {\bibinfo {author} {\bibfnamefont {L.}~\bibnamefont
  {Stebel}}, \bibinfo {author} {\bibfnamefont {M.}~\bibnamefont {Malvestuto}},
  \bibinfo {author} {\bibfnamefont {V.}~\bibnamefont {Capogrosso}}, \bibinfo
  {author} {\bibfnamefont {P.}~\bibnamefont {Sigalotti}}, \bibinfo {author}
  {\bibfnamefont {B.}~\bibnamefont {Ressel}}, \bibinfo {author} {\bibfnamefont
  {F.}~\bibnamefont {Bondino}}, \bibinfo {author} {\bibfnamefont
  {E.}~\bibnamefont {Magnano}}, \bibinfo {author} {\bibfnamefont
  {G.}~\bibnamefont {Cautero}}, \ and\ \bibinfo {author} {\bibfnamefont
  {F.}~\bibnamefont {Parmigiani}},\ }\href {\doibase doi:10.1063/1.3669787}
  {\bibfield  {journal} {\bibinfo  {journal} {Review of Scientific
  Instruments}\ }\textbf {\bibinfo {volume} {82}},\ \bibinfo {pages} {123109}
  (\bibinfo {year} {2011})}\BibitemShut {NoStop}%
\bibitem [{\citenamefont {Krbal}\ \emph {et~al.}(2013)\citenamefont {Krbal},
  \citenamefont {Kolobov}, \citenamefont {Fons}, \citenamefont {Mitrofanov},
  \citenamefont {Tamenori}, \citenamefont {Heged{\"u}s}, \citenamefont
  {Elliott},\ and\ \citenamefont {Tominaga}}]{Krbal:2013vk}%
  \BibitemOpen
  \bibfield  {author} {\bibinfo {author} {\bibfnamefont {M.}~\bibnamefont
  {Krbal}}, \bibinfo {author} {\bibfnamefont {A.}~\bibnamefont {Kolobov}},
  \bibinfo {author} {\bibfnamefont {P.}~\bibnamefont {Fons}}, \bibinfo {author}
  {\bibfnamefont {K.~V.}\ \bibnamefont {Mitrofanov}}, \bibinfo {author}
  {\bibfnamefont {Y.}~\bibnamefont {Tamenori}}, \bibinfo {author}
  {\bibfnamefont {J.}~\bibnamefont {Heged{\"u}s}}, \bibinfo {author}
  {\bibfnamefont {S.~R.}\ \bibnamefont {Elliott}}, \ and\ \bibinfo {author}
  {\bibfnamefont {J.}~\bibnamefont {Tominaga}},\ }\href {\doibase
  10.1063/1.4794870} {\bibfield  {journal} {\bibinfo  {journal} {Applied
  Physics Letters}\ }\textbf {\bibinfo {volume} {102}},\ \bibinfo {pages}
  {111904} (\bibinfo {year} {2013})}\BibitemShut {NoStop}%
\bibitem [{hen(2015)}]{henke:online}%
  \BibitemOpen
  \href@noop {} {}\bibinfo {howpublished}
  {\url{http://henke.lbl.gov/optical_constants/filter2.html}} (\bibinfo {year}
  {accessed December 7, 2015})\BibitemShut {NoStop}%
\bibitem [{\citenamefont {{Raoux, S}}\ and\ \citenamefont {{Wuttig,
  M}}(2010)}]{raoux2010phase}%
  \BibitemOpen
  \bibfield  {author} {\bibinfo {author} {\bibnamefont {{Raoux, S}}}\ and\
  \bibinfo {author} {\bibnamefont {{Wuttig, M}}},\ }\href {\doibase
  10.1007/978-0-387-84874-7} {\emph {\bibinfo {title} {{Phase change
  materials}}}},\ edited by\ \bibinfo {editor} {\bibfnamefont {S.}~\bibnamefont
  {Raoux}}\ and\ \bibinfo {editor} {\bibfnamefont {M.}~\bibnamefont {Wuttig}},\
  science and applications\ (\bibinfo  {publisher} {Springer},\ \bibinfo {year}
  {2010})\BibitemShut {NoStop}%
\bibitem [{\citenamefont {Jund}\ \emph {et~al.}(1997)\citenamefont {Jund},
  \citenamefont {Caprion},\ and\ \citenamefont {Jullien}}]{Jund:1997bz}%
  \BibitemOpen
  \bibfield  {author} {\bibinfo {author} {\bibfnamefont {P.}~\bibnamefont
  {Jund}}, \bibinfo {author} {\bibfnamefont {D.}~\bibnamefont {Caprion}}, \
  and\ \bibinfo {author} {\bibfnamefont {R.}~\bibnamefont {Jullien}},\ }\href
  {http://link.aps.org/doi/10.1103/PhysRevLett.79.91} {\bibfield  {journal}
  {\bibinfo  {journal} {Physical Review Letters}\ }\textbf {\bibinfo {volume}
  {79}},\ \bibinfo {pages} {91} (\bibinfo {year} {1997})}\BibitemShut {NoStop}%
\bibitem [{\citenamefont {Heged{\"u}s}\ and\ \citenamefont
  {Elliott}(2008)}]{Hegedus:2008cn}%
  \BibitemOpen
  \bibfield  {author} {\bibinfo {author} {\bibfnamefont {J.}~\bibnamefont
  {Heged{\"u}s}}\ and\ \bibinfo {author} {\bibfnamefont {S.~R.}\ \bibnamefont
  {Elliott}},\ }\href {\doibase 10.1038/nmat2157} {\bibfield  {journal}
  {\bibinfo  {journal} {Nature Materials}\ }\textbf {\bibinfo {volume} {7}},\
  \bibinfo {pages} {399} (\bibinfo {year} {2008})}\BibitemShut {NoStop}%
\bibitem [{\citenamefont {Rehr}\ \emph {et~al.}(2009)\citenamefont {Rehr},
  \citenamefont {Kas}, \citenamefont {Prange}, \citenamefont {Sorini},
  \citenamefont {Takimoto},\ and\ \citenamefont {Vila}}]{Rehr:2009eu}%
  \BibitemOpen
  \bibfield  {author} {\bibinfo {author} {\bibfnamefont {J.~J.}\ \bibnamefont
  {Rehr}}, \bibinfo {author} {\bibfnamefont {J.~J.}\ \bibnamefont {Kas}},
  \bibinfo {author} {\bibfnamefont {M.~P.}\ \bibnamefont {Prange}}, \bibinfo
  {author} {\bibfnamefont {A.}~\bibnamefont {Sorini}}, \bibinfo {author}
  {\bibfnamefont {Y.}~\bibnamefont {Takimoto}}, \ and\ \bibinfo {author}
  {\bibfnamefont {F.}~\bibnamefont {Vila}},\ }\href {\doibase
  10.1016/j.crhy.2008.08.004} {\bibfield  {journal} {\bibinfo  {journal}
  {Comptes Rendus Physique}\ }\textbf {\bibinfo {volume} {10}},\ \bibinfo
  {pages} {548} (\bibinfo {year} {2009})}\BibitemShut {NoStop}%
\bibitem [{\citenamefont {Rehr}\ \emph {et~al.}(2010)\citenamefont {Rehr},
  \citenamefont {Kas}, \citenamefont {Vila}, \citenamefont {Prange},\ and\
  \citenamefont {Jorissen}}]{Rehr:2010tp}%
  \BibitemOpen
  \bibfield  {author} {\bibinfo {author} {\bibfnamefont {J.~J.}\ \bibnamefont
  {Rehr}}, \bibinfo {author} {\bibfnamefont {J.~J.}\ \bibnamefont {Kas}},
  \bibinfo {author} {\bibfnamefont {F.~D.}\ \bibnamefont {Vila}}, \bibinfo
  {author} {\bibfnamefont {M.~P.}\ \bibnamefont {Prange}}, \ and\ \bibinfo
  {author} {\bibfnamefont {K.}~\bibnamefont {Jorissen}},\ }\href {\doibase
  10.1039/b926434e} {\bibfield  {journal} {\bibinfo  {journal} {Physical
  Chemistry Chemical Physics}\ }\textbf {\bibinfo {volume} {12}},\ \bibinfo
  {pages} {5503} (\bibinfo {year} {2010})}\BibitemShut {NoStop}%
\bibitem [{\citenamefont {Thomsen}\ \emph {et~al.}(1986)\citenamefont
  {Thomsen}, \citenamefont {Grahn}, \citenamefont {Maris},\ and\ \citenamefont
  {Tauc}}]{Thomsen:1986gf}%
  \BibitemOpen
  \bibfield  {author} {\bibinfo {author} {\bibfnamefont {C.}~\bibnamefont
  {Thomsen}}, \bibinfo {author} {\bibfnamefont {H.~T.}\ \bibnamefont {Grahn}},
  \bibinfo {author} {\bibfnamefont {H.~J.}\ \bibnamefont {Maris}}, \ and\
  \bibinfo {author} {\bibfnamefont {J.}~\bibnamefont {Tauc}},\ }\href
  {http://link.aps.org/doi/10.1103/PhysRevB.34.4129} {\bibfield  {journal}
  {\bibinfo  {journal} {Physical Review B}\ }\textbf {\bibinfo {volume} {34}},\
  \bibinfo {pages} {4129} (\bibinfo {year} {1986})}\BibitemShut {NoStop}%
\bibitem [{\citenamefont {Park}\ \emph {et~al.}(2008)\citenamefont {Park},
  \citenamefont {Jung}, \citenamefont {Ryu}, \citenamefont {Choi},
  \citenamefont {Yu}, \citenamefont {Park}, \citenamefont {Han},\ and\
  \citenamefont {Joo}}]{Park:2008dp}%
  \BibitemOpen
  \bibfield  {author} {\bibinfo {author} {\bibfnamefont {I.~M.}\ \bibnamefont
  {Park}}, \bibinfo {author} {\bibfnamefont {J.~K.}\ \bibnamefont {Jung}},
  \bibinfo {author} {\bibfnamefont {S.~O.}\ \bibnamefont {Ryu}}, \bibinfo
  {author} {\bibfnamefont {K.~J.}\ \bibnamefont {Choi}}, \bibinfo {author}
  {\bibfnamefont {B.~G.}\ \bibnamefont {Yu}}, \bibinfo {author} {\bibfnamefont
  {Y.~B.}\ \bibnamefont {Park}}, \bibinfo {author} {\bibfnamefont {S.~M.}\
  \bibnamefont {Han}}, \ and\ \bibinfo {author} {\bibfnamefont {Y.~C.}\
  \bibnamefont {Joo}},\ }\href {\doibase 10.1016/j.tsf.2008.08.194} {\bibfield
  {journal} {\bibinfo  {journal} {Thin Solid Films}\ }\textbf {\bibinfo
  {volume} {517}},\ \bibinfo {pages} {848} (\bibinfo {year}
  {2008})}\BibitemShut {NoStop}%
\bibitem [{\citenamefont {Kalb}\ \emph {et~al.}(2003)\citenamefont {Kalb},
  \citenamefont {Spaepen}, \citenamefont {Leervad~Pedersen},\ and\
  \citenamefont {Wuttig}}]{Kalb:2003vo}%
  \BibitemOpen
  \bibfield  {author} {\bibinfo {author} {\bibfnamefont {J.}~\bibnamefont
  {Kalb}}, \bibinfo {author} {\bibfnamefont {F.}~\bibnamefont {Spaepen}},
  \bibinfo {author} {\bibfnamefont {T.~P.}\ \bibnamefont {Leervad~Pedersen}}, \
  and\ \bibinfo {author} {\bibfnamefont {M.}~\bibnamefont {Wuttig}},\ }\href
  {\doibase 10.1063/1.1610775} {\bibfield  {journal} {\bibinfo  {journal}
  {Journal of Applied Physics}\ }\textbf {\bibinfo {volume} {94}},\ \bibinfo
  {pages} {4908} (\bibinfo {year} {2003})}\BibitemShut {NoStop}%
\bibitem [{\citenamefont {Eesley}(1986)}]{Eesley:1986gk}%
  \BibitemOpen
  \bibfield  {author} {\bibinfo {author} {\bibfnamefont {G.}~\bibnamefont
  {Eesley}},\ }\href {\doibase 10.1103/PhysRevB.33.2144} {\bibfield  {journal}
  {\bibinfo  {journal} {Physical Review B}\ }\textbf {\bibinfo {volume} {33}},\
  \bibinfo {pages} {2144} (\bibinfo {year} {1986})}\BibitemShut {NoStop}%
\bibitem [{\citenamefont {Battaglia}\ \emph {et~al.}(2010)\citenamefont
  {Battaglia}, \citenamefont {Kusiak}, \citenamefont {Schick}, \citenamefont
  {Cappella}, \citenamefont {Wiemer}, \citenamefont {Longo},\ and\
  \citenamefont {Varesi}}]{Battaglia:2010bh}%
  \BibitemOpen
  \bibfield  {author} {\bibinfo {author} {\bibfnamefont {J.~L.}\ \bibnamefont
  {Battaglia}}, \bibinfo {author} {\bibfnamefont {A.}~\bibnamefont {Kusiak}},
  \bibinfo {author} {\bibfnamefont {V.}~\bibnamefont {Schick}}, \bibinfo
  {author} {\bibfnamefont {A.}~\bibnamefont {Cappella}}, \bibinfo {author}
  {\bibfnamefont {C.}~\bibnamefont {Wiemer}}, \bibinfo {author} {\bibfnamefont
  {M.}~\bibnamefont {Longo}}, \ and\ \bibinfo {author} {\bibfnamefont
  {E.}~\bibnamefont {Varesi}},\ }\href {\doibase 10.1063/1.3284084} {\bibfield
  {journal} {\bibinfo  {journal} {Journal of Applied Physics}\ }\textbf
  {\bibinfo {volume} {107}},\ \bibinfo {pages} {044314} (\bibinfo {year}
  {2010})}\BibitemShut {NoStop}%
\bibitem [{\citenamefont {Yamada}\ \emph {et~al.}(1991)\citenamefont {Yamada},
  \citenamefont {Ohno}, \citenamefont {Nishiuchi}, \citenamefont {Akahira},\
  and\ \citenamefont {Takao}}]{Yamada:1991eg}%
  \BibitemOpen
  \bibfield  {author} {\bibinfo {author} {\bibfnamefont {N.}~\bibnamefont
  {Yamada}}, \bibinfo {author} {\bibfnamefont {E.}~\bibnamefont {Ohno}},
  \bibinfo {author} {\bibfnamefont {K.}~\bibnamefont {Nishiuchi}}, \bibinfo
  {author} {\bibfnamefont {N.}~\bibnamefont {Akahira}}, \ and\ \bibinfo
  {author} {\bibfnamefont {M.}~\bibnamefont {Takao}},\ }\href {\doibase
  10.1063/1.348620} {\bibfield  {journal} {\bibinfo  {journal} {Journal of
  Applied Physics}\ }\textbf {\bibinfo {volume} {69}},\ \bibinfo {pages} {2849}
  (\bibinfo {year} {1991})}\BibitemShut {NoStop}%
\end{thebibliography}
\end{document}